\documentclass[12pt]{article}
\usepackage{amssymb}
\usepackage{amsthm}
\usepackage{amsmath}
\usepackage{setspace}
\usepackage{algorithm}
\usepackage{algorithmic}
\usepackage{verbatim}
\usepackage{enumerate}
\usepackage{float}

\makeatletter
\def\inmod#1{\allowbreak\mkern5mu({\operator@font mod}\,\,#1)}
\makeatother

\theoremstyle{definition}

\theoremstyle{remark}

\theoremstyle{definition}

\begin{document}

\medskip

\section{Problem Statement}
In this version of Colored Bin-Packing, we have $n$ items of unit weight, where each item is one of $x$ colors, where $x\geq 3$. Again, we have an unlimited supply of bins, each with weight limit $L$, in which to pack the items and our goal is to minimize the total number of bins.

\section{Algorithm}

The unit-weight case introduces weight constraints. Furthermore, we know that color constraints, specifically the discrepancy, forces us to use more bins than the weight constraint alone. The first half of the algorithm depends on discrepancy. If discrepancy is not a problem (i.e. is not more than 0), we simply order and split the bins up similar to the zero-weight case. If discrepancy is an issue (i.e. is more than 0), we handle the bins differently depending on if bin capacity is even or odd. In both cases we pack bins by alternating between items of the most frequent color and items of other colors. If bin capacity is even, we may be able to condense bins by using items of other colors that top some bins to merge bins containing only a single item of the most frequent color. If bin capacity is odd, we check if we can eventually reduce discrepancy to zero. This is because we use one additional max-color item for every bin we pack. If discrepancy eventually becomes zero, we are in the case where discrepancy is not a problem, and apply that algorithm. If this is not so, we will end up with bins topped by items of the most frequent color. The final set will be optimal, since all the bins will be topped with items of the this color and therefore cannot be combined.

The \textsc{Unit-Weight} algorithm solves this problem.

We first note that we must use at least $\bigg\lceil{\dfrac{n}{L}}\bigg\rceil$ number of bins.

Let $D$ denote the discrepancy, or the difference between the number of $MaxColor$ items and $OtherColors$ items. By the previous proof, we know that for 0-weight items, we will either need one bin (in the case where $D\leq 0$), or $D$ bins (in the case where $D > 0$).

\begin{algorithm}[ht] \caption{Algorithm \textsc{UnitWeight} ($I$). Input is a string $I$ of length $n$, representing items of different colors; $L$ is the weight limit of each bin}
\label{algunit}
\begin{algorithmic}[1]
\STATE Scan $I$ to count the number of times each color occurs. Also keep track of the color that occurs the most; let $MaxColor$ denote this color and let $MaxCount$ denote the number of items of $MaxColor$
\STATE Let $OtherColors$ denote the set of colors that are not $MaxColor$ and let $OtherCount$ denote the number of items of any color in $OtherColors$.
\STATE Let $D = MaxCount - OtherCount$ //Discrepancy
\STATE $S:$ solution //set of bins; initially empty
\IF{$D \le 0$}
	\STATE {\textsc{Split}}($I, n, L$) //\textsc{Split}() gets solution from \textsc{Zero-Weight()} and splits it into bins of capacity $L$
	\STATE Add the solution from {\textsc{Split}} to $S$.
    \STATE \textbf{return} $S$;
\ELSE
	\IF {$L$ is even}
		\STATE Create new current bin $C$. Start with placing a $MaxColor$ item, alternating with an $OtherColors$ item, until C is full. Add $C$ to $S$.
		\STATE Repeat the process until we run out of $OtherColors$ items. Make sure any partially full bins are topped with a $MaxColor$ item. There will be at most one such bin.
		\STATE Place the remaining $MaxColor$ items into bins of their own. Add these bins to $S$.
		\STATE Call  \textsc{Condense($S$)} to condense all the single-item bins. Add these bins to $S$.
	\ELSIF {$L$ is odd}
    \IF {$D \leqslant   \lceil \frac{OTHER}{\lfloor{L/2}\rfloor} \rceil $}
    \STATE //if there are enough OtherColors items such that alternating will allow the discrepancy to be reduced to zero

				\FOR{1 to $D$}
					\STATE Create new current bin $C$. Start with placing a $MaxColor$ item, alternating with an $OtherColors$ item, until C is full. Add $C$ to $S$.
				\ENDFOR
				\STATE Call {\textsc{Split}}($R$), where $R$ is the remaining items to be placed. //D=0 now so \textsc{Split} may be called
				\STATE Add the solution from {\textsc{Split}} to $S$.
			\ELSE
				\STATE Create new current bin $C$. Start with placing a $MaxColor$ item, alternating with an $OtherColors$ item, until C is full. Add $C$ to $S$.
				\STATE Repeat this until all $OtherColors$ items are used. Make sure any partially full bins are topped with a $MaxColor$ item. There will be at most one such bin.
				\STATE Place the remaining $MaxColor$ items in bins of their own. Add these to $S$.
			\ENDIF
	\ENDIF

\ENDIF
\end{algorithmic}
\end{algorithm}

\begin{algorithm}
\hspace{-1in}
\caption{Algorithm \textsc{Split} ($I, n, L)$}
\begin{algorithmic}[1]
	\STATE $R$ = \textsc{Zero-Weight}($I$):
	\FOR {$i = 1$ to $\bigg\lceil{\dfrac{n}{L}}\bigg\rceil$}
        \STATE Create new bin $C$
		\STATE Remove the first $L$ items (or remaining items if there are fewer than $L$) from $R$ and pack into $C$
        \STATE Add $C$ to $S$
	\ENDFOR
\end{algorithmic}
\end{algorithm}

\begin{algorithm}\caption{Algorithm \textsc{Condense} ($S$). $S$ is a set of bins, both completely and partially filled}
\begin{algorithmic}[1]

	\STATE Separate $S$ into three sets respectively called M-bins, P-bins and F-bins.
\STATE M-bins (MaxColor bins): contain only one item and the item is of MaxColor.
\STATE P-bins (partially full bins): contain both $MaxColor$ and $OtherColors$ items, are topped with a MaxColor item, and can fit at least two more items. There will be at most one of this bin type.
\STATE F-bins (full bins): fully packed bins that are topped with an $OtherColors$ item.
\STATE //Note that since \textsc{Condense} is called only when $L$ is even, all fully packed bins (F-bins) will be topped with an $OtherColors$ item, therefore all bins fall into one of these three categories.

        \IF {P-bins is not empty}
        		\STATE Let $current$ be the bin in P-bins
                \STATE Remove $current$ from P-bins
		\ELSIF {M-bins is not empty}
		\STATE Let $current$ be a bin in M-bins
		\STATE Remove $current$ from M-bins
	\ENDIF
	\WHILE {F-bins is not empty and M-bins is not empty}
		\IF {weight of $current$ $+ 2 > L$}
			\STATE Remove $current$ from its containing set
            \STATE Let $current$ be the next bin in M-bins  //switch to a new bin since $current$ is full
		\ENDIF
		\STATE Let $x$ be the top-most $OtherColors$ item from a bin in F-bins
        \STATE Let $y$ be a $MaxColor$ item from a bin (or the bin??) in M-bins
        \STATE Pack $x$ followed by $y$ into $current$.
		\STATE Remove $x$ and $y$ from their containing bins.
	\ENDWHILE
\end{algorithmic}
\end{algorithm}

\newpage

\section{Proof}
\textbf{Case 1:} $D \leqslant 0$ (first if statement) If the discrepancy is not more than zero, we use the \textsc{Zero-Weight} algorithm to first order the items so they satisfy the color requirements. Then, without breaking this order, we pack the first $L$ items in the first bin, the next $L$ items in the second, and so on, until we have packed all the items.
\\
\\
\\
\\
\noindent \textbf{Example:}\\
\underline{Items:} 4 W, 3 B, 2 Y\\
\underline{Weight Limit:} 3\\
In the zero-weight case, we would sort these items in this order, using only one bin:\\
\begin{center}BWBWBWYWY\\
\end{center}
We use this result to get the unit weight solution:\\
\begin{center}BWB WBW YWY\\
\end{center}
where a space denotes a new bin.
\\
\\
This is based on the two requirements defined by the problem: color and weight. We know from the zero-weight case that the \textsc{Zero-Weight} algorithm satisfies the color restriction when discrepancy is less than or equal to 0. By allocating the items according to the weight limit, we use the minimum amount of bins necessary based on bin capacity. Since both are satisfied, the solution is optimal.
\\
\\
\\
\textbf{Case 2:} $D > 0$ (else statement)
\begin {enumerate}
\item {When $L$ is even.}

\noindent \textbf{Example:}\\
\underline{Items:} 12 W, 3 B, 2 Y 2 G\\
\underline{Weight Limit:} 4\\
\underline{D:}  12 - (3 + 2 + 2) = 12 - 7 = 5\\
We use  {\textsc{Unitweight}} to first pack these into bins,\\
\begin{center} WBWB  WYWG  WBWY  WGW  W  W  W  W\\
\end{center}
where a space denotes a new bin. Note that there are four bins that only contains a single  $MAXCOLOR$ - white - item.
We reduce the number of bins further by calling {\textsc{Condense}}.\\
This becomes,
\begin{center}WBW  WYW  WBWY  WGW  WBW  WGW\\
\end{center}
giving us 6 bins instead of 8.
\\
\\
Note that we use the same definition of N-bins, H-bins and F-bins seen in {\textsc{Condense}}.

With a positive discrepancy, it is easy to see that the color requirement will force us to have more bins than the minimum number of bins enforced by the weight limit. This is because each excess $MaxColors$ item requires an individual bin - these bins form the set of `M-bins'.  \\
	
Keeping this in mind, we also want to use $OtherColors$ items to buffer as many $MaxColor$ items as possible. We do this to minimize the effect excess $MaxColor$ items will have on the eventual number of bins. The more $MaxColor$ items we buffer, the fewer M-bins we are forced to create. Hence we start every bin with a $MaxColor$ item, and place only one $OtherColors$ item between each $MaxColor$ item. Completely packed bins topped with an $OtherColors$ item form the set called 'F-bins'. \\

When we run out of $OtherColors$ items while packing a bin, we get a potential P-bins. To use up as many $MaxColor$ items as possible, we cover the $OtherColors$ item currently topping the bin, with a $MaxColor$ item. If the bin is able to accept 2 or more items after we do so, we get a P-bins. Because this happens only when we run out of $OtherColors$ items, a P-bins only occurs once.\\

We refer to the packing before the call to \textsc{Condense} as the \textit{initial packing}. After the initial packing, $S$ contains F-bins, M-bins and a possible P-bins. Because $L$ is even and all F-bins start with a $MaxColor$ item, all F-bins must be topped with an $OtherColors$ item. These $OtherColors$ items are used to combine the M-bins, and further reduce our total number of bins using the algorithm {\textsc{Condense}}. \\

To show {\textsc{Condense}}'s optimality, we prove that the number of bins in $S$ cannot be further reduced after {\textsc{Condense}} is run.\\\\
There are several cases, all of which are dependent on {\textsc{Condense}}'s exit condition.

\begin{enumerate}[i.]
\item $S$ contains no F-bins.\\

\item $S$ contains no M-bins.\\
%
%

\item $S$ contains neither F-bins nor M-bins.
\end{enumerate}
Since combining bins requires both F-bins and M-bins, any one of these three conditions means that we cannot reduce the total number of bins by combining any bins in $S$.
\item {When $L$ is odd.}

We observe that when the weight limit is odd, if we use as many $MaxColor$ items as possible when packing a bin, we use one more $MaxColor$ than $OtherColors$ item for every bin. To do so, we fill the bin in a similar fashion to if $L$ is even - starting with a $MaxColor$ item and alternating with $OtherColors$ items. However, because the $L$ is odd, we end with a $MaxColor$ item. This is the excess $MaxColor$ item we pack when packing each bin.\\

Next, we note that if $D > 0$, the number of M-bins depends on the number of $OtherColors$ items. This is because since we alternate between $OtherColors$ and $MaxColor$ items when packing the bins, each bin takes exactly  $\lfloor {L/2} \rfloor$ $OtherColors$ items. If $OtherCount$ is the total number of $OtherColors$, we can form $\lceil \frac{OtherCount}{\lfloor{L/2}\rfloor} \rceil$ P-bins and F-bins. The remaining $MaxColor$ items must each be packed in a separate bin which are the M-bins.\\

Thus, if we can form enough P-bins and F-bins to pack the excess $MaxColor$ items, we can reduce $D$ to 0, otherwise we form M-bins. This gives us 2 cases.

\begin{enumerate}
\item{We eventually reduce $D$ to 0.}

\noindent \textbf{Example:}\\
\underline{Items:} 8 W, 3 B, 2 Y 2 G\\
\underline{Weight Limit:} 5\\
\underline{D:}  8 - (3 + 2 + 2) = 8 - 7 = 1\\
We use {\textsc{Unitweight}} to first pack these into bins.\\
After the first bin,
\begin{center} WBWBW\\
\end{center}
where a space denotes a new bin, {\textsc{Discrepancy}} is now zero and we pack the remaining items using {\textsc{Split}}.
We first sort the remaining items,\\
\begin{center}WYWGWBWYWG\\
\end{center}
before splitting the sorted items into bins to give us,
 \begin{center}WYWGW  BWYWG\\
\end{center}
This gives us,
\begin{center}WBWBW  WYWGW  BWYWG\\
\end{center}
as our overall solution.
\\
\\

If $D \leqslant   \lceil \frac{OtherCount}{\lfloor{L/2}\rfloor} \rceil $, we know that $D = 0$ after $D$ bins are filled in Step -- (fill in). From Case 1, we know that {\textsc{Split}} gives us an optimal packing when $D = 0$. Thus, for $S$ to be optimal, the $D$ bins packed during this phase must also be optimal.

To show that the $D$ bins are optimally packed, we note that the initial packing falls into two categories:\\

\begin{enumerate}
\item {Initial packing yields a P-bin:}

If the initial packing yields a P-bin, it means that $D = 1$ when we started filling the P-bin. The phase ends when we place the last excess $MaxColor$ item into the P-bin. This reduces $D$ to 0.\\

In this case, we show that there are no more items left. \\

Since $D = 0$, we know the number of $MaxColor$ and $OtherColors$ items must be equal. Any additional $OtherColors$ item would mean the P-bins would not exist, as the last item placed into the bin would be an $OtherColors$ item. This is a contradiction. Since there are no $OtherColors$ items remaining there must be no $MaxColor$ items remaining. Since there are no items remaining,  we cannot combine any of the bins, and the solution set is optimal.\\

\item {Initial packing does not yield any P-bins.}

Note that we will not have any M-bins. When $D=0$, there are two cases:

\begin{enumerate}
\item $MaxCount = OtherCount = 0$:

We won't need any additional bins so clearly there are no M-bins.

\item $MaxCount = OtherCount > 0$:
Since there are no excess $MaxColor$ items, there will be no bins containing only a $MaxColor$ item.

Here, the solution set contains only F-bins. Since we have no M-bins, bin combination cannot happen, and the set is optimal. \\
 \end{enumerate}

 \item{$D$ remains greater than 0.}

\noindent \textbf{Example:}\\
\underline{Items:} 15 W, 3 B, 2 Y 2 G\\
\underline{Weight Limit:} 5\\
\underline{D:}  15 - (3 + 2 + 2) = 15 - 7 = 6\\
We use {\textsc{Unitweight}} to pack these into bins, knowing that $D$ will not eventually reduce to zero.
\begin{center} WBWBW  WYWGW  WBWYW  WGW  W  W  W  W\\
\end{center}
where a space denotes a new bin. Though there are bins containing only one items, we cannot condense them since all the non-$MAXCOLOR$ items need them in their bins. The solution is optimal.\\
\\
\\

This means we did not have enough P-bins and/or F-bins to pack excess $MaxColor$ items and therefore cannot reduced $D$ to 0. $S$ contains at least one M-bins. However, because all the bins are topped by $MaxColor$ items, we cannot combine any bins to reduce the total number of bins. $S$ is optimal.
\end{enumerate}
\end{enumerate}

\end{enumerate}

\section{Time Complexity}
We analyze the time complexity based on the various cases.
If $D$ is less than or equal to 0, the algorithm first calls \textsc{Split} which runs \textsc{Zero-Weight} to order the items based on color, then packs the items based on weight, using at $O(n)$ time in total where $n$ is the number of items.

If $D$ is more than 0, we consider two sub-cases: $L$ is odd or $L$ is even.
If $L$ is odd, we check if we can eventually reduce the $D$ to 0. If we can, we pack items until $D$ is 0. This takes at most $O(n)$, since we consider each item only
once. When $D$ is 0 we call \textsc{Split} which takes $O(n)$ time.  If we cannot
reduce discrepancy to 0, we pack items by alternating colors. Since each item is considered exactly once, this takes $O(n)$ time.

If $L$ is even, again we pack by alternating colors then call \textsc{Condense}. The runtime of \textsc{Condense} depends on the
number of bins and since there will be $O(n)$ bins, \textsc{Condense} runs in $O(n)$ time.

\bigskip

\section{Bin Coloring Problem with 0-weight items of more than 2 colors, reordering allowed}

In this section, we consider a version of the Bin Coloring Problem where items are of more than 2 colors, all items have a weight of 0, and reordering is allowed. More formally, the problem is defined as follows:

\begin{itemize}
\item The input to the problem is a string. Each character $x$ represents an item with the color $x$.
\item All items have a weight of 0.
\item No adjacent items in the bin can be of the same color.
\item Reordering is allowed. For example, for $i$ and $j$ such that $i>j$, the $i$th item in the string can be packed either before or after the $j$th item in the string.
\end{itemize}

We provide an algorithm \textsc{zero-weight} that is optimal for this problem.

Since the only constraint in the 0-weight situation is the color constraint (that no two adjacent items can have the same color), we pay special attention to the most frequent color, i.e. the color with the most number of items. In the first case, if this color has no more items than all the other colors combined, we can pack all the items in one bin, using other color items as buffers between items of the most frequent color. Specifically, we first alternate between items of other colors until the we have exactly enough items of other colors to buffer the items of the most frequent color. Then, we alternate between items of other colors and items of the most frequent color. If on the other hand, the most frequent color occurs more than all the other colors combined, we are forced to use all the other color items as buffers. Even so, we will be running out of the other color items in the end and left with several items of the most frequent because there are too many of them. For each of the items of the  most frequent color that are remaining, we are forced to open one new bin, since they are all of the same color.

\begin{algorithm}\caption{Algorithm \textsc{zero-weight}. Input is a string $I$, representing items of different colors. }
\label{algo1}
\begin{algorithmic} [1]
\STATE Scan $I$ to count the number of times each color occurs. Also keep track of the color that occurs the most; let $MaxColor$ denote this color and let $MaxCount$ denote the number of items of $MaxColor$
\STATE Let $OtherColors$ denote the set of colors that are not $MaxColor$ and let $OtherCount$ denote the number of items of any color in $OtherColors$.
\STATE $Discrepancy = MaxCount - OtherCount$
\IF {$Discrepancy \le 0$}
\STATE Create a new bin $B$ whose size is equal to the length of $I$.
\STATE Alternate between an item of $MaxColor$ and an item of any arbitrary $OtherColors$ until $B$ is full and all items are packed. Only one bin is needed in this case.
\STATE \textbf{return} $B$
\ELSE
\STATE Create an empty set $S$ that will be the solution
\STATE Create a bin $B_0$ of size $OtherCount * 2 + 1$. Add $B_0$ to $S$.
\STATE Alternate between an item of $MaxColor$ and an item of any arbitrary $OtherColors$ until $B_0$ is full. When $B_0$ is full, there will be no items of $OtherColors$ left.
\STATE Create a bin $B_x$ for each of the remaining $x$ items of $MaxColor$ and pack each of these items in one of these bins. In this case, the total number of bins is $Discrepancy$. Add each $B_x$ to $S$
\STATE \textbf{return} $S$
\ENDIF
\end{algorithmic}
\end{algorithm}

\begin{proof}

First note that from a previous paper (Bin Packing with colors), if there are only two colors, then the minimum number of bins necessary is the discrepancy of the two colors. For example, if there are 5 Black items and 3 White items, the number of bins necessary is $5-3=2$ bins. This is done by starting with an item of the color with more items ($B$) and alternating with items of the other color ($W$) until $W$ runs out. We then add one more item of color $B$ into this first bin. If $|B$ and $|W|$ denote the number of black and white items, respectively, we need $|B|-|W|-1$ more bins to fit the rest of the items of color $W$ for a total of $|B|-|W|$ bins.

For more than 2 colors, we consider two cases:

\underline{Case 1}: $Discrepancy \le 0$:
See below for an example.
In this case there are at least as many $OtherColors$ items as there are $MaxColor$ items. In this case, we need only one bin, which is the fewest number of bins any algorithm can get. To do this, we alternate between items of $OtherColors$ until the sum of the unused $OtherColors$ items is one less than the number of $MaxColors$ items. We then add one $MaxColor$ item to the bin. From here, we alternate between any $OtherColors$ item and a $MaxColor$ item until all items are used. The last item will be one of $MaxColor$.

Note that it is always possible to alternate the $OtherColors$ items until $OtherCount$ is one less than the number of $MaxColor$ items. The only case when we can no longer alternate is when there are items of only one color, $X$, remaining, and because of the color restriction, we can no longer put these items into the bin. We claim that at this point, $OtherCount$, which in this case is the number of $X$ items, will be one less than $MaxCount$. We prove this by way of contradiction. Suppose when we stop alternating, $OtherCount$ is equal or greater than $MaxCount$. Note that an $X$ item must be on top of the bin, otherwise we have not completed the alternating step since we can pack another $X$ item on top of the bin. There must be at least $MaxCount$ $X$ items remaining, so there were at least $MaxCount + 1$ total $X$ items, which contradicts our assumption that $MaxColor$ is the color with the maximum number of items.

After these alternations, we have one more $MaxColor$ items than all $OtherColors$ items, and it is easy to see that we can alternate between them and put all items in the same bin.

\underline{Case 2}: $Discrepancy > 0 $. In this case, all other items can be regarded as some uniform color \textit{non-MaxColor} that is not equal to $MaxColor$, which makes this the black/white bin packing problem. This is because there are so many $MaxColor$ items that we must pack each \textit{non-MaxColor} item between two $MaxColor$ items to use as few bins as possible. The first bin $B_0$ will start with a \textit{non-MaxColor} item, alternate $MaxColor$ and \textit{non-MaxColor} items, and end with a \textit{non-MaxColor} item, so will have $OtherCount$*2+1 items in total. Each additional bin will have an item of $MaxColor$. Then the minimum number of bins necessary is still the difference between the number of $MaxColor$ and \textit{non-MaxColor} items, i.e. the discrepancy, as is shown in the Black/White bin packing problem.

\end{proof}

The first step of the algorithm requires a scan of the input string to determine $OtherCount$ and $MaxCount$. The second step requires a scan to separate the items into in two sets: items of $OtherColors$ and items of $MaxColor$.  Packing each item takes constant time, so the remainder of the algorithm runs linear to the number of items. Therefore the algorithm has total runtime $O(3n)$ where $n$ is the number of items.

\bigskip
\bigskip

\noindent \textsc {Example: }

\medskip

\noindent \underline {Items:} WWWWWWWWBBYY (8W, 2B, 2Y)

\medskip

\noindent In this case, $MaxColor$ is $W$. $MaxCount$ equals $8$ and $OtherCount$ equals $4$. $Discrepancy > 0$.

\medskip

\noindent According to our algorithm, we first create a bin $B_0$ of size $4*2+1=9$. Then we alternate between $W$ and any $OtherColor$ to fill the first bin.

\medskip

\begin{center} WBWBWYWYW
\end{center}

\medskip

\noindent Now, we have used all the $OtherColors$, and we still have 3 $W$'s remaining. Therefore, we create $3$ new bins, each holding one of the remaining $W$'s.

\medskip

\noindent Therefore, we get the following final result:

\medskip

\begin{center} WBWBWYWYW / W / W / W
\end{center}

\end{document}